\documentclass[12pt]{article}
\pdfoutput=1

\usepackage{color}
\usepackage{graphicx}
\usepackage{amsmath}
\usepackage{slashed}
\usepackage{amssymb}
\usepackage{epstopdf}

\input xy
\xyoption{all}
\xyoption{web}

\usepackage[
      colorlinks=true,
      linkcolor=blue,
      urlcolor=blue,
      filecolor=black,
      citecolor=red,
      pdfstartview=FitV,
      pdftitle={},
        pdfauthor={Michael Gutperle, Yi Li},
        pdfsubject={},
        pdfkeywords={},
        pdfpagemode={},
        bookmarksopen=true
      ]{hyperref}

\usepackage{color}
\usepackage{graphicx}

\usepackage{sectsty}
\sectionfont{\large}

\renewcommand{\thefootnote}{\fnsymbol{footnote}}
\renewcommand{\thanks}[1]{\footnote{#1}}
\newcommand{\starttext}{
\setcounter{footnote}{0}
\renewcommand{\thefootnote}{\arabic{footnote}}}

\newcommand{\bea}{\begin{eqnarray}}
\newcommand{\eea}{\end{eqnarray}}
\newcommand{\be}{\begin{equation}}
\newcommand{\ee}{\end{equation}}

 \newcommand{\reals}{\mathbb{R}}

\def\half{ {1\over 2}}

\DeclareMathOperator{\tr}{tr}
\usepackage{ dsfont } 

\long\def\symbolfootnote[#1]#2{\begingroup%
\def\thefootnote{\fnsymbol{footnote}}\footnote[#1]{#2}\endgroup}

\begin{document}
\setlength{\baselineskip}{18pt}

\starttext
\setcounter{footnote}{0}

%
\bigskip

\begin{center}

{\Large \bf  Higher Spin Lifshitz Theory and Integrable Systems}

\vskip 0.4in

{\large  Michael Gutperle and Yi Li}

\vskip .2in

{ \it Department of Physics and Astronomy }\\
{\it University of California, Los Angeles, CA 90095, USA}\\[0.5cm]
\href{mailto:yli@physics.ucla.edu}{\texttt{yli@physics.ucla.edu}}\texttt{, }\href{mailto:gutperle@physics.ucla.edu}{\texttt{gutperle@physics.ucla.edu}}

\bigskip

\bigskip

\end{center}

\begin{abstract}

\setlength{\baselineskip}{18pt}

In this note we construct asymptotically Lifshitz spacetimes in the Chern-Simons formulation of three dimensional higher spin gravity and relate the resulting theories to integrable systems which are elements of the KdV hierarchy.

\end{abstract}

\setcounter{equation}{0}
\setcounter{footnote}{0}

%
%
%
%
%
\newpage


\section{Introduction}
\setcounter{equation}{0}
\label{sec1}
Higher spin gravities in various dimensions as formulated by Vasiliev and collaborators (see \cite{Vasiliev:2000rn,Vasiliev:2004qz,Bekaert:2005vh,Didenko:2014dwa} for some reviews) provide an interesting new playground to explore the AdS/CFT correspondence. In the following we will consider only three dimensional higher spin theory, which can be formulated using Chern-Simons gauge theories \cite{Blencowe:1988gj,Bergshoeff:1989ns}.
Interest in these theories is fueled by the proposal of an exact  AdS/CFT duality linking such theories to  $W_N$ minimal model CFTs, due to Gaberdiel and Gopakumar \cite{Gaberdiel:2010pz,Gaberdiel:2012uj}. 

Three dimensional higher spin theories allow for the construction of non-AdS  solutions \cite{Gary:2012ms,Afshar:2012nk,Gonzalez:2013oaa,Gary:2014mca},  such as asymptotically Lobachevsky, Schr\" odinger, warped AdS and Lifshitz spacetimes. In the following we will focus on the asymptotically Lifshitz solutions following the approach developed in  \cite{Gutperle:2013oxa}. On the field theory side systems with Lifshitz scaling, i.e.  anisotropic scaling symmetries  with respect to spatial and time directions, are ubiquitous and important in condensed matter systems near  quantum critical points (see e.g. \cite{Kachru:2008yh}). 

The goal of the present paper is  to construct an asymptotically Lifshitz spacetime  using the Chern-Simons formulation with various integer values of the Lifshitz scaling exponent  $z$ and uncover the relation of these spacetimes to integrable systems, in particular members of the KdV hierarchy.

The structure of this paper is as follows: In section \ref{sec2} we give a brief overview  of the 
Chern-Simons formulation of three-dimensional higher spin gravity (more details can be 
found  for example in \cite{Gaberdiel:2012uj,Ammon:2012wc,Campoleoni:2012hp,Henneaux:2010xg}). In section 
\ref{sec3} we review some aspects of the integrable systems which are relevant for the present paper.  In particular, we discuss the formulation of 
the KdV hierarchy in terms of pseudo differential operators and the Lax pair formulation. 
This formalism will be useful to make the connection to the asymptotically Lifshitz solutions 
of  the Chern-Simons higher spin theory. In section \ref{sec4} we briefly review  the general properties of field theories with Lifshitz symmetry. In section \ref{sec5} we construct the general algorithm to find  asymptotic Lifshitz connections in the context of gauge algebra $hs(\lambda)$ with an integer scaling exponent $z
$. We work out the detailed calculation for the special case of Lifshitz exponent $z=3$ for  the  $hs(\lambda)$ algebra. Finally, the infinite dimensional set of equations is reduced to a finite dimensional set by setting $\lambda=4$, for which $hs(\lambda)$ is 
truncated to $sl(4,\reals)$, and we obtain a $z=3$ Lifshitz spacetime for  $sl(4,\reals)$ Chern-Simons gravity.  In section \ref {sec6} the map between the 
$sl(4,\reals),z=3$ and $sl(3,\reals),z=2$ Lifshitz theories and members of  KdV hierarchy is presented. A specific gauge choice,  called KdV gauge, must be made  to  make the relation work. A conjecture on a general relation valid for all values of $z$ and $sl(N,\reals)$  as well as for the infinite dimensional $hs(\lambda)$ case, is given. In 
section \ref{sec7}  the symmetry algebra for asymptotic Lifshitz connections in the $hs(\lambda)$ theory is   constructed for arbitrary $z$. In addition  the two specific cases $sl(4,\reals),z=3$ and $sl(3,\reals),z=2$  are worked out in KdV gauge. We close the paper  in section \ref{sec8}  with a discussion of some open questions and future directions of research. Our conventions for the relevant gauge algebras and gauge choices are presented in appendix \ref{appa} and \ref{appb} respectively.

\section{Review of Chern-Simon formulation of higher spin theories}
\setcounter{equation}{0}
\label{sec2}

The Chern-Simons formulation of three dimensional (higher-spin) gravity is based
on two copies of
Chern-Simons action at level $k$ and $-k$ and gauge algebra $sl(N,\reals)\times
sl(N,\reals)$ or the higher spin algebra $hs(\lambda)\times hs(\lambda)$. For 
completeness we present our conventions for the algebras in appendix  \ref{appa}.
The action is given by,

\be\label{chernsimonsa}
S=S_{CS}[A]-S_{CS}[\bar A]
\ee
where the Chern-Simons action takes the familiar form,
\be
S_{CS}[A]= {k\over 4\pi}  \int {\rm tr}\Big( A\wedge dA+{2\over 3} A\wedge A\wedge
A\Big)
\ee
The equations of motion impose the flatness of the gauge connection
\be
F=dA+ A\wedge A=0, \quad \quad \bar F=d\bar A+ \bar A\wedge \bar A=0
\ee
It was shown in \cite{Witten:1988hc,Achucarro:1987vz} that 
Einstein gravity with negative cosmological constant is realized by choosing  the gauge algebra  to be $sl(2,\reals)\times sl(2,\reals)$.  The relation to the metric is obtained by expressing the vielbein and spin connection in terms of the Chern-Simons connection as follows
\be\label{vielbein}
e_\mu ={l\over 2} (A_\mu -\bar A_\mu ), \quad \omega_\mu=\half (A_\mu +\bar
A_\mu)
\ee
The metric can be calculated from the connections via
\be\label{metrica}
g_{\mu\nu} = {1\over 2} tr (e_\mu e_{\nu})
\ee
 The gauge transformations of the  Chern-Simons theory 
  \be\label{gauget}
\delta A = d\Lambda + [A, \Lambda], \quad \delta \bar A = d\bar \Lambda + [A, \bar
\Lambda]
\ee
correspond to diffeomorphisms and Lorentz frame rotations  in the metric theory.
For $N>2$ the theory is a truncation of the three dimensional Vasiliev theory to fields of spin $s=2,3,\cdots,N$.
The simplest case is $N=3$  which corresponds to the gauge algebra $sl(3,\reals)\times sl(3,\reals)$  was discussed in \cite{Campoleoni:2010zq},  where it was shown that the theory described gravity coupled to a massless spin three field which is given in terms of the gauge connection (\ref{vielbein}) by
\be
 \phi_{\mu\nu \rho} ={1\over
6} \tr(e_{(\mu}
e_\nu
e_{\rho)})
\ee
analogous formulae can be obtained for larger $N$ following \cite{Campoleoni:2011hg}, but will not  be needed here.
The analysis of the asymptotic symmetry is greatly simplified by  special choice of gauge. We define a  radial coordinate
$\rho$, where the
holographic boundary will be located at $\rho\to \infty$. In addition we define a
time-like coordinate $t$ and a space-like coordinate $x$, which can be either compact or non-compact
and hence the
boundary
has either the topology of $\reals\times S^1$ or $\reals\times \reals$.  The
``radial gauge" which
we
will use is
constructed by defining $b= \exp( \rho L_0)$ and expressing the radial dependence via a $\rho$ dependent gauge transformation
\be\label{bigadef}
A_\mu(x,t,\rho) = b^{-1} a_\mu(x,t)  \,b + b^{-1}\partial_\mu b, \quad \bar A_\mu(x,t,\rho) = b \,\bar
a_\mu(x,t) b^{-1} + b
\,\partial_\mu  (b^{-1})
\ee
Here $L_0$ is a Cartan generator of a $sl(2,\reals)$ sub-algebra of $sl(N,\reals)$, or its corresponding generator $V^2_0$ in $hs(\lambda)$. In this "radial gauge" the flatness condition reduces to a condition on the $\rho$ independent connections $a_t$  and $a_x$.
\be\label{flat}
\partial_t a_x -\partial_x a_t + [a_t,a_x]=0, \quad \partial_t \bar{a}_x -\partial_x \bar{a}_t + [\bar{a}_t,\bar{a}_x]=0
\ee
The Chern-Simons formulation of three dimensional  higher spin theories has been used to define black holes in such theories via holonomy conditions \cite{Gutperle:2011kf,Ammon:2011nk,  Perez:2012cf,Perez:2013xi}, as well as to calculate entanglement entropies using  Wilson-loops   \cite{deBoer:2013vca,Ammon:2013hba,Castro:2014mza}.

\section{Review of  the KdV hierarchy}
\setcounter{equation}{0}
\label{sec3}
The KdV equation  is a partial differential equation describing propagation of (shallow) water waves in channels, given by 
\be\label{kdveqa}
4{ \partial u\over \partial t}=  {\partial^3 u\over \partial x^3}+ 6 u {\partial u\over \partial x}
\ee
The KdV equation is an example of an integrable system, with infinitely many conserved and commuting 
charges,  as well as  soliton solutions with dispersion-free scattering. The KdV equation is a 
particular example of an infinite set of integrable systems, the so called KdV hierarchy. We 
review  the treatment of the KdV hierarchy using pseudo-differential operators and Lax 
pairs.  Pseudo differential operators allow for the introduction of negative powers of derivatives $\partial$ retaining the rules of differentiation,   such as the Leibniz rule. More information about the formalism and its applications to integrable systems can be found in   \cite{dickey,battle92}.

The KdV hierarchy is characterized by two integers 
$n$ and $m$ and a differential operator $L$
\be
L= \partial^n+ u_2 \partial^{n-2} +\cdots +u_{n-1} \partial + u_n
\ee
Here $\partial= {\partial \over \partial x}$ and $u_i=u_i(x,t)$. The formalism of pseudo differential operators allows to define  fractional powers of $L$, in particular $L^{1/n}$.
\be
L^{1/ n} = \partial + {1\over n} u \partial^{-1}+ o(\partial^{-2})
\ee
For another integer $m$ one defines
\be
P_m = \Big(L^{m/n}\Big)_+
\ee
Where the subscript $()_+$ denotes the non-negative part of the pseudo differential operator, which has terms with $\partial^k, k\geq 0$. An integrable system is constructed due to the fact that $P,L$ form a Lax pair, i.e. the evolution equation

\be\label{laxpair}
{\partial\over \partial t}L=[P_m ,L] 
\ee 
gives a system of partial differential equations  for $u_i(x,t)$.
For the KdV hierarchy an infinite set of conserved quantities can be obtained by 
\be\label{conserved}
q^{(k)} = \int {\rm res} \big(L^{k\over n} \big)
\ee
Where ``${\rm  res}$" denotes the coefficient of the term multiplying  $\partial^{-1}$ in the 
pseudo differential operator. The charges are conserved if the equation of motion 
(\ref{laxpair}) is satisfied and the fields $u_i$ fall off fast enough as $x\to \pm \infty$, so 
that total derivatives can be discarded.
In the following we will present several members of the KdV hierarchy for low $m$ and $n$, 
for which we will show that they are related to Lifshitz higher spin theories.

\subsection{KdV equation: $n=2,m=3$}
The original KdV equation, fits in the hierarchy by choosing $n=2$ and $m=3$. The Lax pair  is given by
\be
L=\partial^2 +u_2
\ee
and 
\bea
P_3&=& \Big( L^{3\over 2}\big)_+\nonumber\\
&=&  \partial^3 + {3\over 2} u_2 \partial + {3\over 4} u_2' 
\eea
The commutator gives
\be
[P_3,L]= {1\over 4} u_2'''+ {3\over 2} u_2 u_2'
\ee
The Lax equation (\ref{laxpair}) takes the following form 
\be 
4\dot u_2 =u_2'''+ 6 u_2 u_2'
\ee
which reproduces the KdV equation (\ref{kdveqa}).

\subsection{Boussinesq equation: $n=3,m=2$}\label{subsec32}
The next case is given by choosing $n=3$ and $m=2$. As we shall see later this case will be relevant for the $z=2$ Lifshitz. The operator $L$  is now of third order and contains two independent fields $u_2$ and $u_3$
\be
L= \partial^3 + u_2 \partial + u_3
\ee
and
\be
L^{1/3}= \partial + {1\over 3} u_2 \partial^{-1} + \frac{1}{3}(u_3-u_2') \partial^{-2} + o(\partial^{-3})
\ee
Setting $m=2$ the Lax operator becomes
\be
P_2 = (L^{2/3})_+= \partial^2+ {2\over 3} u_2
\ee 
The Lax equation (\ref{laxpair}) is equivalent to the following system of partial differential equations
\bea\label{boussia}
\dot u_2 &=& 2 u_3' - u_2'' \nonumber \\
\dot u_3 &=& u_3''-{2\over 3} u_2'''-{2\over 3}  u_2 u_2'
\eea
Eliminating $u_3$ then gives an equation for  $u_2$ alone
\be
\ddot{u}_2= -{1\over 3} u_2'''-{4\over 3} (u_2 u_2')'
\ee
This equation  is known as the Boussinesq equation \cite{boussinesqref}, which has been studied in the context of propagation of waves.  
Using (\ref{conserved}) we can easily calculate the first two  conserved charges
\bea
q^{(1)}&=& \int {\rm res} \big( L^{1 \over 3} \big) = {1 \over 3} \int u_2 \nonumber\\
q^{(2)}&=& \int {\rm res} \big( L^{2 \over 3} \big) = \int (\frac{2}{3}u_3 - \frac{1}{3}u_2') = \frac{2}{3} \int u_3 
\eea

\subsection{$n=4,m=3$ member of  KdV hierarchy }

The case $n=4$ $m=3$ will be relevant for the $z=3$ Lifshitz theory. The operator $L$ is now of fourth order and contains three fields $u_i$, $i={2,3,4}$.
\be
L= \partial^4 + u_2 \partial^2+ u_3 \partial + u_4
\ee
from this we can evaluate 
\be
 L^{1/4}=\partial+\dfrac{u_2}{4}\partial^{-1}+\dfrac{1}{4}(u_3-\dfrac{3}{2}u_2^{'})\partial^{-2}+(\frac{1}{4}u_4-\frac{3}{8}u_3^{'}+\frac{5}{16}u_2^{''}-\frac{3}{32}u_2^2) \partial^{-3} + o(\partial^{-4})
\ee
The Lax operator is given by

\be
 P_3=L^{3/4}_+=\partial^3+\dfrac{3}{4}u_2\partial+\dfrac{3}{4}u_3-\dfrac{3}{8}u_2^{'}
 \ee
The Lax equation (\ref{laxpair}) is equivalent to
\bea
  \dot{u_2}&=&\dfrac{1}{4}u_2^{'''}-\dfrac{3}{2}u_3^{''}+3u_4^{'}-\dfrac{3}{4}u_2u_2^{'} \nonumber\\
\dot{u_3}&=&-2u_3^{'''}+3u_4^{''}+\dfrac{3}{4}u_2^{''''}-\dfrac{3}{4}u_2u_3^{'}-\dfrac{3}{4}u_3u_2^{'} \nonumber\\
 \dot{u_4}&=&u_4^{'''}+\dfrac{3}{8}u_2^{'''''}-\dfrac{3}{4}u_3^{''''}+\dfrac{3}{4}u_2u_4^{'}\nonumber\\
&& -  \dfrac{3}{4}u_2u_3^{''}+\dfrac{3}{8}u_2u_2^{'''}-\dfrac{3}{4}u_3u_3^{'}+\dfrac{3}{8}u_3u_2^{''}. 
\eea
Using  (\ref{conserved}) we can calculate the conserved quantities and display the first three here
\bea
q^{(1)}&=& \int {\rm res} \big( L^{1 \over 4} \big) = {1 \over 4} \int u_2 \nonumber\\
q^{(2)}&=& \int {\rm res} \big( L^{2 \over 4} \big) = \int (\frac{1}{2}u_3 - \frac{1}{2}u_2') = \frac{1}{2} \int u_3 \nonumber\\
q^{(3)}&=& \int {\rm res} \big( L^{3 \over 4} \big) = \int (\frac{3}{4}u_4-\frac{3}{8}u_3'+\frac{1}{16}u_2''-\frac{3}{32}u_2^2) =\int (\frac{3}{4}u_4-\frac{3}{32}u_2^2) 
\eea

\section{Field theories with Lifshitz scaling}
\setcounter{equation}{0}
\label{sec4}

Lifshitz theories are field theories which exhibit   an  anisotropic scaling symmetry with
respect to space and
time
\be\label{lifscale}
t\to \lambda^z t, \qquad x\to \lambda x
\ee
where $z$ is the Lifshitz  scaling exponent and $z=1$ corresponds to conformal scaling. The algebra of  Lifshitz symmetries is given by the time
translation $H$, 
the spatial
translation $P$ and the Lifshitz scaling $D$, which satisfy the following commutation relations
\be
  [P,H]=0 \qquad
  [D,H]=z H \qquad
  [D,P]= P \label{lifalg}
\ee
For theories with Lifshitz scaling the stress tensor does not have to be
symmetric, as there is no relativistic boost symmetry. The  stress energy  tensor contains four components:  the energy density
${\cal E}$, the
energy flux ${\cal E}^x$, the momentum density ${\cal P}_x$ and the stress density
$\Pi_x^{\;x}$.
These quantities satisfy the following conservation equations (see e.g.
\cite{Ross:2011gu}).
\bea\label{emcoma}
\partial_t {\cal E}+ \partial_x {\cal E}^x&=&0 \nonumber \\
\partial_t {\cal P}_x+ \partial_x {\Pi}_x^{\; x}&=&0
\eea
In addition, the Lifshitz scaling with exponent $z$ implies a modified
tracelessness condition
\be\label{emcomb}
z {\cal E}+ {\Pi}_x^{\; x}=0.
\ee
Since the operator $D$ generates scale transformation  the commutation relations (\ref{lifalg}) imply that the momentum operator $P$ has scaling dimension one, whereas the Hamiltonian $H$ has scaling dimension $z$. We will give the precise definition of scaling dimension later.

It is an interesting question whether theories with Lifshitz scaling have a holographic description. 
For a two dimensional Lifshitz theory the three dimensional spacetime takes the following form:
\be\label{lifmetb}
 ds^2=d\rho^2-e^{2z\rho}dt^2+e^{2\rho}dx^2
\ee
The shift $\rho \to \rho + \ln \lambda$ in the holographic radial coordinate  induces a Lifshitz scaling transformation on $t,x$ with scaling exponent $z$ as in (\ref{lifscale}). Such 
metrics (and their higher dimensional generalizations) are not solutions of Einstein gravity 
and nontrivial matter interactions have to be added. The first solutions of this kind where 
found in  \cite{Kachru:2008yh} in four dimensional gravity coupled to anti-symmetric tensor 
fields. Subsequently, Many   solutions which exhibit Lifshitz asymptotics have been 
constructed in supergravity theories.

\section{Asymptotic Lifshitz connection}
\setcounter{equation}{0}
\label{sec5}
In this section we construct connections which are asymptotically Lifshitz in  the Chern-Simons formulation of three dimensional higher spin gravity. 
A Lifshitz connection is a solution to flatness condition which produces a  Lifshitz metric (\ref{lifmetb}). 
The connections which reproduce a metric with integer  scaling exponent $z$ can be easily written down using the  $hs(\lambda)$ algebra, the unbarred connection is given by
\bea\label{hslama}
  a &=&  V^{z+1}_z dt + V^2_1 dx, \quad \quad A = V^{z+1}_z e^{z\rho} dt + V^2_1 e^{\rho} dx + V^2_0 d\rho
  \eea
and the barred connection is given by
\bea\label{hslamb}
  \bar{a} &=& V^{z+1}_{-z} dt + V^2_{-1} dx, \quad \quad \bar{A} = V^{z+1}_{-z} e^{z\rho} dt +V^2_{-1} e^{\rho} dx - V^2_0 d\rho
\eea
It follows from (\ref{metrica}) that these connections realize  a Lifshitz spacetime with an arbitrary integer $z$. For an integer $\lambda=N$ the algebra $hs(\lambda)$ is truncated to $sl(N,\reals)$. For example (\ref{hslama}) in the  $z=2$ case, one reproduces the $sl(3,\reals)$ Lifshitz connections studied in \cite{Gutperle:2013oxa} with the identification $V^3_{ \pm 2}=W_{\pm 2}, V^2_{\pm 1}=L_{\pm 1}$ and $V^2_0=L_0$.

 In the following we will  consider connections where the barred sector is determined in terms of the unbarred sector. This is possible due to   an automorphism of $hs(\lambda)$ 
 algebra, which is obtained from  a conjugation $(V^s_m)^c=(-1)^{s+m+1}V^s_{-m}$. In 
 particular the generator $V^2_0$ used in constructing the radial gauge transformations is 
 self conjugate up to a sign, i.e.  $(V^2_0)^c=-V^2_0$. Consequently, if  $A$ solves the 
 flatness condition $F=0$ in the radial gauge,  the barred connection is chosen to be the  
 conjugate  $\bar{A}=A^c$. For this choice $\bar A$ automatically satisfies the flatness 
 condition   $\bar{F}=0$ and the radial gauge. From now on 
 we will leave out the barred sector as it is determined from the un-barred sector.

Though we have explicit expression for Lifshitz connections, they are static solutions without any dynamics. Here we want to consider asymptotic Lifshitz connection. Asymptotic Lifshitz connections are connections in which leading terms are Lifshitz connections given by (\ref{hslama}) where additional terms are present with sub-leading powers $e^\rho$. Consequently such connections will lead to asymptotic Lifshitz spacetimes where the metric and tensor fields have additional terms which become negligible as $\rho\to \infty$ compared to the Lifshitz vacuum.

\subsection{Constructing asymptotic Lifshitz connections}

In this section we describe an algorithm to find solutions to the flatness conditions  in the radial gauge 
\be\label{flatd}
\partial_t a_x - \partial_x a_t + [a_t,a_x] = 0
\ee
for asymptotically Lifshitz $hs(\lambda)$ connections. 
However the   implementation of the algorithm becomes unwieldy for larger values of $z$ and hence a detailed calculation is presented for the case $z=3$ in section \ref{sec52}.
Furthermore, if we set $\lambda=N$ the Lie algebra $hs(\lambda)$ is truncated to $sl(N,\reals)$ and a finite dimensional example is treated in section \ref{sec54}.

\medskip

The algorithm proceeds in the following steps:
\begin{enumerate}
\item Adopt a ``lowest weight gauge"  for $a_x$  such that it only contains lowest weight terms except for $V^2_1$. Lowest weight terms are of the form $V^i_{-i+1}$, whose weight is lowest for a given spin.
\be
a_x = V^2_1 + \sum_{i=2}^\infty \alpha_i V^i_{-i+1} 
\ee
Where $\alpha_{i}(x,t)$ are the dynamical fields and their evolution equations will be determined.
\item The ansatz for the time component of the connection is given by
\be\label{atans}
a_t = (*a_x)^z |_{traceless} + \Delta a_t
\ee
Where the subscript $\mid _{traceless}$ denotes the removal of the $V^1_0$ component from the star product, see appendix \ref{hslamconv} for our conventions on the higher spin algebra.
This  ansatz works for  any integer value of $z$. The ansatz is motivated by the fact that for constant $a_x$, eq.(\ref{atans}) is a solution of the flatness condition with $\Delta a_t=0$\footnote{Note that this construction has been used to construct higher spin black holes \cite{Kraus:2012uf}.}. 

\item For $x$ and $t$ dependent $\alpha_i$ the flatness condition (\ref{flatd}) now takes the form
\be\label{faltz}
\partial_t a_x -\partial_x ( *a_x)^z |_{traceless} -\partial_x \Delta a_t + [\Delta a_t,a_x]=0
\ee
\item Calculate $( *a_x)^z |_{traceless}$ and note that the terms with the highest weight which appear in this expression are $V^i_{-i+2z-1}$ for sufficiently large spin $i$. Considering the only term that raises the weight is the commutator with $V^2_1$ in $a_x$, it is sufficient to include terms with weight up to $-i+2z-2$ for $\Delta a_t$ in general, so the whole expression will be closed of terms with weight up to $-i+2z-1$. In addition, we can add a term $V^i_{-i+2z-1}$ to $\Delta a_t$ if it happens to be a highest weight term, which has a vanishing commutator with $V^2_1$. Now we have the suitable ansatz for $a_t$.
\item Since $\partial_t a_x$ only contains lowest weight term, so all the non lowest weight terms in $-\partial_x ( *a_x)^z |_{traceless} -\partial_x \Delta a_t + [\Delta a_t,a_x]$ must vanish. This will fix most coefficients in $\Delta a_t$ in terms of coefficients in $a_x$.
\item Require lowest weight terms $V^i_{-i+1}$ to cancel out in the flatness condition we get the time evolution of the $\alpha_i$.
\end{enumerate}

\subsection{Detailed calculation for the $z=3$ case}
\label{sec52}
In this section  we would like to work out the explicit example of $z=3$ to illustrate the general algorithm. The ansatz for $\Delta a_t$ is
\be
\Delta a_t = \sum_{i=2}^\infty \beta_i V^i_{-i+1} + \sum_{i=2}^\infty \gamma_i V^i_{-i+2} + \sum_{i=2}^\infty \delta_i V^i_{-i+3} + \sum_{i=3}^\infty \sigma_i V^i_{-i+4} + \mu V^3_2.
\ee
In the following we will calculate each term in the flatness condition (\ref{faltz}). The triple product
\bea
a_x*a_x*a_x &=& (V^2_1 + \sum_{i=2}^\infty \alpha_i V^i_{-i+1})^3 \nonumber\\
&=& (V^2_1)^3 + \sum_{i=2}^\infty \alpha_i (V^2_1*V^2_1*V^i_{-i+1} + V^2_1*V^i_{-i+1}*V^2_1 + V^i_{-i+1}*V^2_1*V^2_1) \nonumber\\
&+& \sum_{i=2}^\infty \sum_{j=2}^\infty \alpha_i \alpha_j (V^2_1*V^i_{-i+1}*V^j_{-j+1} + V^i_{-i+1}*V^j_{-j+1}*V^2_1 + V^j_{-j+1}*V^2_1*V^i_{-i+1}) \nonumber\\
&+& \sum_{i=2}^\infty \sum_{j=2}^\infty \sum_{k=2}^\infty \alpha_i \alpha_j \alpha_k V^i_{-i+1}*V^j_{-j+1}*V^k_{-k+1}.
\eea
The relevant products are evaluated as follows 
\be
V^2_1*V^2_1*V^i_{-i+1} + V^2_1*V^i_{-i+1}*V^2_1 + V^i_{-i+1}*V^2_1*V^2_1 =o(i,\lambda)V^{i-2}_{-i+3} + p(i,\lambda)V^i_{-i+3} + 3 V^{i+2}_{-i+3} 
\ee
and
\bea
&&V^2_1*V^i_{-i+1}*V^j_{-j+1} + V^i_{-i+1}*V^j_{-j+1}*V^2_1 + V^j_{-j+1}*V^2_1*V^i_{-i+1} \nonumber\\
&& \; \; = r(i,j,\lambda)V^{i+j-2}_{-i-j+3} + s(i,j)V^{i+j-1}_{-i-j+3} + 3 V^{i+j}_{-i-j+3} 
\eea
as well as
\be
V^i_{-i+1}*V^j_{-j+1}*V^k_{-k+1} = V^{i+j+k-2}_{-i-j-k+3}
\ee
where $o(i,\lambda)$, $p(i,\lambda)$, $r(i,j,\lambda)$ and $s(i,j)$ are coefficients which can be calculated from structure constants defined in appendix \ref{hslamconv}. The commutator
\be
[\Delta a_t,a_x] = [\sum_{i=2}^\infty \beta_i V^i_{-i+1} + \sum_{i=2}^\infty \gamma_i V^i_{-i+2} + \sum_{i=2}^\infty \delta_i V^i_{-i+3} + \sum_{i=3}^\infty \sigma_i V^i_{-i+4} + \mu V^3_2,V^2_1 + \sum_{i=2}^\infty \alpha_j V^j_{-j+1}]
\ee
can be calculated using the   commutation relation of the generators of the algebra
\be
[V^2_1,V^t_n]=(t-n-1)V^t_{n+1}
\ee
and
\bea
&&[V^i_{-i+1},V^j_{-j+1}]=0 \nonumber\\
&&[V^i_{-i+2},V^j_{-j+1}]=(j-1)V^{i+j-2}_{-i-j+3} \nonumber\\
&&[V^i_{-i+3},V^j_{-j+1}]=2(j-1)V^{i+j-2}_{-i-j+4} \nonumber\\
&&[V^i_{-i+4},V^j_{-j+1}]=w(i,j,\lambda)V^{i+j-4}_{-i-j+5} + 3(j-1)V^{i+j-2}_{-i-j+5} \nonumber\\
&&[V^i_{-i+5},V^j_{-j+1}]=z(i,j,\lambda)V^{i+j-4}_{-i-j+6} + 4(j-1)V^{i+j-2}_{-i-j+6}.
\eea
where $w(i,j,\lambda)$ and $z(i,j,\lambda)$ can be calculated from structure constants.
After some simplification the flatness condition now reads
\bea
 \sum_{i=2}^\infty \dot{\alpha_i} V^i_{-i+1} &=& \sum_{i=2}^\infty \alpha_{i+2}^{'} o(i+2,\lambda) V^i_{-i+1} + \sum_{i=2}^\infty \alpha_i^{'} p(i,\lambda) V^i_{-i+3} + \sum_{i=4}^\infty \alpha_{i-2}^{'} 3V^i_{-i+5} \nonumber\\
&& + \sum_{i=2}^\infty \sum_{j=2}^i (\alpha_{i-j+2} \alpha_j)^{'} r(i-j+2,j,\lambda) V^i_{-i+1} + \sum_{i=4}^\infty \sum_{j=2}^{i-2} (\alpha_{i-j} \alpha_j)^{'} 3V^i_{-i+3}\nonumber\\
&& + \sum_{i=4}^\infty \sum_{j+k+l=i+2} (\alpha_j\alpha_k\alpha_l)^{'} V^i_{-i+1}  + \sum_{i=2}^\infty \beta_i^{'} V^i_{-i+1} + \sum_{i=2}^\infty \gamma_i^{'} V^i_{-i+2} + \sum_{i=2}^\infty \delta_i^{'} V^i_{-i+3}  \nonumber\\
&&+ \sum_{i=3}^\infty \sigma_i^{'} V^i_{-i+4}+ \mu^{'} V^3_2 + \sum_{i=2}^\infty (2i-2)\beta_i V^i_{-i+2}  + \sum_{i=2}^\infty (2i-3)\gamma_i V^i_{-i+3} \nonumber\\
&&+ \sum_{i=2}^\infty (2i-4)\delta_i V^i_{-i+4} + \sum_{i=3}^\infty (2i-5)\sigma_i V^i_{-i+5} - \sum_{i=2}^\infty \sum_{j=2}^i \gamma_{i-j+2} \alpha_j (j-1) V^i_{-i+1} \nonumber\\
&&- \sum_{i=2}^\infty \sum_{j=2}^i \delta_{i-j+2} \alpha_j 2(j-1) V^i_{-i+2} - \sum_{i=3}^\infty \sum_{j=2}^{i-1} \sigma_{i-j+2} \alpha_j 3(j-1) V^i_{-i+3} \nonumber\\
&& - \sum_{i=2}^\infty \sum_{j=2}^{i+1} \sigma_{i-j+4} \alpha_j w(i-j+4,j,\lambda) V^i_{-i+1} - \sum_{i=2}^\infty \mu \alpha_{i+1} z(3,i+1,\lambda) V^i_{-i+2} \nonumber\\
&& - \sum_{i=3}^\infty 4(i-2) \mu \alpha_{i-1} V^i_{-i+4}.\label{flatnesstbt}
\eea 
Vanishing of non lowest weight terms enables us to solve most coefficients in $a_t$ in terms of the coefficients in $a_x$, i.e. $\alpha_i$'s.
From equating the  coefficients of the $V^i_{-i+5}$ generators we  find
\bea
\sigma_3&=&0 \nonumber\\
\sigma_i&=&-{3 \over 2i-5}\alpha_{i-2}^{'},i \geq 4
\eea
Equating the  coefficients of the $V^i_{-i+4}$ generators gives
\bea
\delta_3&=&0 \nonumber\\
\delta_i&=&{3 \over (2i-5)(2i-4)}\alpha_{i-2}^{''}, i \geq 4
\eea
Equating the  coefficients of the $V^i_{-i+3}$ generators gives
\bea
\gamma_2&=&-p(2,\lambda)\alpha_2^{'}-\delta_2^{'} \nonumber\\
\gamma_3&=&-\frac{1}{3}p(3,\lambda)\alpha_3^{'} \nonumber\\
\gamma_i&=&-\frac{1}{2i-3}p(i,\lambda)\alpha_i^{'}-\frac{1}{(2i-3)(2i-4)(2i-5)}\alpha_{i-2}^{'''}\nonumber\\
&& -\frac{3}{2i-3}\sum_{j=2}^{i-2}(2+{3(j-1)\over 2(i-j)-1})\alpha_j \alpha_{i-j}^{'}, i \geq 4 
\eea
Finally equating the  coefficients of the $V^i_{-i+2}$ generators  gives

\bea
\beta_2&=&\frac{1}{2}p(2,\lambda)\alpha_2^{''}+\frac{1}{2}\delta_2^{''}+\alpha_2\delta_2 \nonumber\\
\beta_3&=&\frac{1}{12}p(3,\lambda)\alpha_3^{''}+\delta_2\alpha_3 \nonumber\\
\beta_i&=&\frac{1}{(2i-2)(2i-3)}p(i,\lambda)\alpha_i^{''}+\frac{1}{(2i-2)(2i-3)(2i-4)(2i-5)}\alpha_{i-2}^{''''} \nonumber\\
&& +\frac{3}{(2i-2)(2i-3)}\sum_{j=2}^{i-2}(2+\frac{3(j-1)}{2(i-j)-1})(\alpha_j \alpha_{i-j}^{'})^{'}+\alpha_j\delta_2\nonumber\\
&& +\frac{3}{2i-2}\sum_{j=2}^{i-2}\frac{j-1}{(i-j)(2(i-j)-2)}\alpha_j \alpha_{i-j}^{''}, i\geq 4
\eea
There are two exceptions, $\delta_2$ and $\mu$ cannot be determined by equations of motion (\ref{flatnesstbt}) and can in principle be chosen arbitrarily. It can be shown that they are purely gauge.
Vanishing of lowest weight terms gives us equations of motion
\bea
&&\dot{\alpha_i}-o(i+2,\lambda)\alpha_{i+2}^{'}-\sum_{j=2}^i r(i-j+2,j,\lambda)(\alpha_{i-j+2}\alpha_j)^{'}-\sum_{j+k+l=i+2}(\alpha_j\alpha_k\alpha_l)^{'}-\beta_i^{'} \nonumber\\
&&+\sum_{j=2}^i (j-1)\gamma_{i-j+2}\alpha_j+\sum_{j=2}^{i+1} w(i-j+4,j,\lambda)\sigma_{i-j+4}\alpha_j=0
\eea \label{EOM}
where the cubic term is understood to be there only for $i\geq4$. We get the equations of motion in terms of $\alpha$'s with extra gauge freedom to choose $\delta_2$ and $\mu$ arbitrarily.

\subsection{Scaling dimension}\label{sec53}

An interesting feature of the equations of motion is their scaling structure. For concreteness we will work out the scaling structure for  the   $z=3$ Lifshitz  example. The results can easily be adapted for general integer values of $z$. A scaling transformation (\ref{lifscale})   acts on the space and time coordinates $x\rightarrow \lambda x$ $t\rightarrow \lambda ^3 t$. We show in the following that  the connections are   invariant after  an appropriate rescaling of the fields.
A field variable has scaling dimension $l$ if it is rescaled by a factor $\lambda^{-l}$.

In the triple product  $a_x*a_x*a_x$ there are terms of the form $\alpha_i\alpha_j V^2_1*V^i_{-i+1}*V^j_{-j+1}$ as well as  terms  of the form $\alpha_{i+j}V^2_1*V^2_1*V^{i+j}_{-i-j+1}$. Both kind of terms contain higher spin generators of  of weight $3-i-j$ and hence should have the same scaling dimensions, hence the dimensions of $\alpha_i
\alpha_j$ and $\alpha_{i+j}$ should be the same, or symbolically $[\alpha_i]+
[\alpha_j]=[\alpha_{i+j}]$, where the square bracket means the scaling dimension of the quantity. 
Consequently,  the scaling dimension of $\alpha_i$ is additive and determined by the weight of the generator $V^i_{-i+1}$.

Following the same argument, comparing the triple product of $a_x$ to $\Delta a_t$ gives the relation $[\alpha_i]=[\beta_{i-2}]=[\gamma_{i-1}]=[\delta_i]=[\sigma_{i+1}]$ and $
[\mu]=[\sigma_2]$. Comparing  $\partial_x a_t$ to $[a_t,a_x]$ terms we get $[\partial_x]
[\beta_{i-1}]=[\beta_i]$. Comparing  $\partial_t a_x$ to $[a_t,a_x]$ terms gives  $[\partial_t]
[\alpha_{i-3}]=[\alpha_i]$.  It follows from (\ref{lifscale})  that $[\partial_x]=1$ and $
[\partial_t]=3$. Summarizing the dimensions of all field variables
\be\label{scalinga}
[\alpha_i]=i,\; [\beta_i]=i+2, \;  [\gamma_i]=i+1, \;  [\delta_i]=i, \; [\sigma_i]=i-1, \; [\mu]=1,
\ee
The scaling dimensions (\ref{scalinga}) agree with the scaling of the fluctuating fields demanding the invariance of the connections  $A_t dt$ and $A_x dx$  under a shift of the radial coordinate $\rho$ accompanied by a Lifshitz scaling of $x,t$ given by
\bea
\rho^{'}&=&\rho+\log \lambda  \nonumber\\
 x^{'}&=&\lambda^{-1} x \nonumber\\
 t^{'}&=&\lambda^{-3} t
\eea
 
 To fix the gauge and to get equations of motion in terms of $\alpha$'s unambiguously we want to express $\delta_2$ and $\mu$ in terms of $\alpha$'s or their derivatives. $\mu$ has to be zero since it's the only variable with dimension 1. $\delta_2$ must be proportional to $\alpha_2$, say, with proportionality constant $c$.

\subsection{The  $sl(4,\reals)$ and $z=3$ example}
\label{sec54}
Following the algorithm described above, we can in principle construct asymptotic Lifshitz connections with infinitely many terms in the context of $hs(\lambda)$. 
In this section we simplify further and consider a finite truncation of the infinite dimensional algebra.  We set $\lambda=4$ so the higher spin algebra is truncated to $sl(4,\reals)$. In the equations of motion (\ref{EOM}) $\dot{\alpha}_i$ is coupled to $\alpha$'s with higher scaling dimension only via the second term. $o(5,\lambda)$ and $o(6,\lambda)$ are zero when $\lambda=4$, because they must vanish due to their definition as coefficients of spin five and spin six $hs(\lambda)$ elements in (5.6). Even though there are still  infinitely many $\alpha$'s, the dynamics of the first three $\alpha_2,\alpha_3,\alpha_4$ are decoupled from the others and it is consistent to set all the $\alpha_i, i>4$ to zero. Consequently  when the algebra is truncated, the equations of motion are also truncated to what we will get if we just start with ansatz in the finite algebra. As discussed in section \ref{sec53} the scaling symmetry imposes  the gauge choices   $\mu=0$ and  $\delta_2=c\alpha_2$. Now we have
\bea
&& \sigma_3=0, \;\; \sigma_4=-\alpha_2^{'} \nonumber\\
&& \delta_2, \delta_3=0, \;\;  \delta_4=\dfrac{1}{4}\alpha_2^{''} \nonumber\\
&& \gamma_2=\dfrac{41}{5}\alpha_2^{'}-\delta_2^{'}, \;\; \gamma_3=2\alpha_3^{'}, \;\; \gamma_4=-\dfrac{1}{20}\alpha_2^{'''}-\dfrac{3}{5}\alpha_4^{'}-\dfrac{9}{5}\alpha_2\alpha_2^{'} \nonumber\\
&& \beta_2=\delta_2\alpha_2+\dfrac{1}{2}\delta_2^{''}-\dfrac{41}{10}\alpha_2^{''},\;\;   \beta_3=\delta_2\alpha_3-\dfrac{1}{2} \alpha_3^{''}, \nonumber\\
&& \beta_4=\delta_2\alpha_4+\dfrac{1}{120}\alpha_2^{''''}+\dfrac{1}{10}\alpha_4^{''}+\dfrac{3}{10}(\alpha_2^{'})^2+\dfrac{23}{60}\alpha_2\alpha_2^{''}
\eea
and the equations of motion
\bea
 \dot{\alpha_2}&=&-(\dfrac{41}{10}-\dfrac{1}{2}c)\alpha_2^{'''}-(\dfrac{123}{5}-3c)\alpha_2^{'}\alpha_2+\dfrac{54}{5}\alpha_4^{'} \nonumber\\
 \dot{\alpha_3}&=&-\dfrac{1}{2}\alpha_3^{'''}-(15-c)\alpha_3^{'}\alpha_2-(30-3c)\alpha_2^{'}\alpha_3 \nonumber\\
 \dot{\alpha_4}&=&\dfrac{1}{10}\alpha_4^{'''}+\dfrac{1}{120}\alpha_2^{'''''}-(30-4c)\alpha_2^{'}\alpha_4-(\dfrac{27}{5}-c)\alpha_2\alpha_4^{'}-12\alpha_3^{'}\alpha_3+\dfrac{13}{30}\alpha_2\alpha_2^{'''} \nonumber\\
&&+ \dfrac{59}{60}\alpha_2^{'}\alpha_2^{''}+\dfrac{24}{5}\alpha_2^2\alpha_2^{'}
\eea \label{EOM43}

\section{Map to KdV}
\setcounter{equation}{0}
\label{sec6}
In this section we want to demonstrate that there is a map from the equations of motion  of the Chern-Simons connection to the time  evolution equations of the KdV hierarchy. We discuss two concrete examples $z=2$, $sl(3,\reals)$ and the $z=3$, $sl(4,\reals)$ and then  propose a conjecture for the general case.

\subsection{The $sl(3,\reals)$ and $z=2$ example}

Here we we want to consider a simpler case which has been worked out in the previous paper \cite{Gutperle:2013oxa} and provide the map to the $n=3,m=2$ member of the KdV hierarchy\footnote{A different realization of the Boussinesq integrable system in Chern-Simons higher spin gravity was presented in \cite{Compere:2013gja}.} which was described in section \ref{subsec32}.
The asymptotic Lifshitz connection is\footnote{Here we have adapted the general notation, with $-\mathcal{L}$ in the previous paper \cite{Gutperle:2013oxa} replaced by $\alpha_2$ and $\mathcal W$ replaced by $\alpha_3$.}
\bea
  a_t &=& V^3_2 + 2\alpha_2 V^3_0 - \frac{2}{3}\alpha_2' V^3_{-1} - 2\alpha_3
  V^2_{-1} +
  \Big(\alpha_2^2 +\frac{1}{6}\alpha_2''\Big)V^3_{-2} \label{cfalt}\\
  a_x &=& V^2_1 + \alpha_2 V^2_{-1} + \alpha_3 V^3_{-2} \label{cfalx}
\eea
The flatness condition  implies the equations of motion
\bea\label{evolut}
  \dot{\alpha_2} &=& -2\alpha_3'\\
  \dot{\alpha_3} &=& \frac{4}{3} (\alpha_2^2)' + \frac{1}{6}\alpha_2'''
\eea
These equations are equivalent to the Boussinesq equations (\ref{boussia}) via the following field identification
\bea
 u_2&=&4 \alpha_2 \nonumber\\
 u_3&=&-4 \alpha_3 + 2 \alpha_2^{'} 
\eea
or conversely
\bea
\alpha_2&=&\dfrac{1}{4}u_2 \nonumber\\
\alpha_3&=&-\dfrac{1}{4}u_3+\dfrac{1}{8}u_2^{'}.
\eea

\subsection{The $sl(4,\reals)$ and $z=3$ example}
\label{subsec62}

Let's go back to the $sl(4,\reals), z=3$ case, which has the novelty of gauge dependence described by the parameter $c$. We want to find a map from Chern-Simons connection variables to KdV variables such that the equations of motion of Chern-Simons connection (\ref{EOM43}) are equivalent to KdV with $n=4, m=3$ (3.19). The Chern-Simons variables have scaling dimensions according to the analysis in the section \ref{sec53}. The  KdV variables also have scaling dimensions by the formulation of pseudo-differential operators. The fact that the scaling dimensions on both sides have to agree puts strong restrictions on the mapping of the variables. Hence we must use the ansatz $u_2=k\alpha_2,u_3=a\alpha_2^{'}+b\alpha_3$. For  the second KdV equation 
\be
\dot{u_3}=-2u_3^{'''}+3u_4^{''}+\dfrac{3}{4}u_2^{''''}-\dfrac{3}{4}u_2u_3^{'}-\dfrac{3}{4}u_3u_2^{'}\ee
on the right hand side $\alpha_2\alpha_3^{'}$ and $\alpha_2^{'}\alpha_3$ have the same 
coefficients, on the left hand side the same kind of terms come from $\dot{\alpha_3}=-
\dfrac{1}{2}\alpha_3^{'''}-(15-c)\alpha_3^{'}\alpha_2-(30-3c)\alpha_2^{'}\alpha_3$, so we 
must have $(15-c)=(30-3c)$, obtaining $c=\dfrac{15}{2}$. Comparing terms and check  integrability condition recursively one can obtain $k=10,a=10,b=24$ and the full map
\bea
&& u_2=10\alpha_2 \nonumber\\
&& u_3=10\alpha_2^{'}+24\alpha_3 \nonumber\\
&& u_4=3 \alpha_2^{''}+9\alpha_2^2+12\alpha_3^{'}+36\alpha_4
\eea
establish the correspondence.

Here we see we must make the gauge choice $c=\frac{15}{2}$ to establish the relation between Chern-Simons Lifshitz theory and KdV hierarchy. We call this kind of gauge choice ``KdV gauge".
Explicitly, the equations of motion in the KdV gauge read
\bea
&& \dot{\alpha_2}=-\dfrac{7}{20}\alpha_2^{'''}-\dfrac{21}{10}\alpha_2^{'}\alpha_2+\dfrac{54}{5}\alpha_4^{'} \nonumber\\
&& \dot{\alpha_3}=-\dfrac{1}{2}\alpha_3^{'''}-\frac{15}{2}\alpha_3^{'}\alpha_2-\frac{15}{2}\alpha_2^{'}\alpha_3 \nonumber\\
&& \dot{\alpha_4}=\dfrac{1}{10}\alpha_4^{'''}+\dfrac{1}{120}\alpha_2^{'''''}+\dfrac{21}{10}\alpha_2\alpha_4^{'}-12\alpha_3^{'}\alpha_3+\dfrac{13}{30}\alpha_2\alpha_2^{'''}+\dfrac{59}{60}\alpha_2^{'}\alpha_2^{''}+\dfrac{24}{5}\alpha_2^2\alpha_2^{'}
\eea

\subsection{General conjecture}

 In the two previous sections we have mapped the equations of motion for asymptotic Lifshitz connections to member of the  KdV hierarchy in two particular cases, namely  the $z=2$, $sl(3,\reals)$ connection is mapped to the $n=3, m=2$ element of the  KdV  hierarchy  and  $z=3$, $sl(4,\reals)$  connection is mapped  to the  $n=4,m=3$ element of the  KdV  hierarchy. This result inspires us to propose general conjecture:
 
 \medskip
 
 \noindent  The asymptotic Lifshitz connection for $sl(N,\reals)$ and an arbitrary integer Lifshitz scaling exponent $z$ can be mapped to the member of the KdV hierarchy with $n=N, m=z$.  
 
 \medskip

Apart from the two cases worked out in this paper we have also checked the case $z=2$, $sl(4,\reals)$. The fact that $n=N$ can be deduced from the fact that lowest weight ansatz for $a_x$ for $sl(N,\reals)$ contains $N-1$ fields $a_i$ which has to be equated with the $n-1$ fields $u_i$ on the KdV side. Furthermore, the dimensional analysis for the Lax equation (\ref{laxpair}) implies $\partial_t$ has the same dimension as $P_m$, that is, the same dimension as $\partial_x^m$, so $m$ is exactly the Lifshitz scaling exponent.

Now a natural question rises for  the Lifshitz connection in the algebra $hs(\lambda)$ where $\lambda$ is not an integer:  Can the equations of motion, which involve infinite number of fields be mapped  to some integrable hierarchy?  Note that such an integrable system should reduce to the KdV hierarchy when $hs(\lambda)$ is truncated to  $sl(N,\reals)$ upon setting $\lambda=N$. A candidate for such  integrable systems is the  KP hierarchy which we briefly review here.

The starting point is the following pseudo differential operator which contains infinitely many fields $v_i, i=2,3,\cdots$.
\be
S=\partial+v_2\partial^{-1}+v_3\partial^{-2}+v_4\partial^{-3}+\ldots
 \ee
 The Lax equation for the $m$-th element of the hierarchy\footnote{Note that the name ``KP hierarchy" is usually reserved for the system of equations for all $m$ where a different time variable $t_m$ is associated with each element. We are interested in the a specific element of the hierarchy and denote the time simply by $t$.} 
   is defined by
 \be
 {\partial \over \partial t} S=[S^m_+,S] 
\ee
The Lax equation gives equations of motion of the KP variables $v$'s.

The connection of the KP hierarchy to  the  KdV hierarchy is obtained as follows:  Note that the Lax equation above implies the following equation for the $n$-th power of the operator $S$
\be
 \dot{S^n}=[S^m_+,S^n] 
\ee
With the definitions $L=S^n$ and $P_m= S^{m\over n}_+$ we get the Lax equation of  KdV defined in (\ref{laxpair}).  At this point the pseudo differential operator $L$ contains all possible 
powers of $\partial$, down to $\partial^{-\infty}$. It is possible to consistently restrict $L$ to   only non-negative powers of differentiation, which  implies that  the dynamics of the first 
$n-1$ variables is decoupled from the other, and they are just KdV hierarchy with the same values of $m$ and $n$. Consequently, it is possible to perform  a field redefinition to 
truncate KP to KdV. The map from $sl(N,\reals), z$ Chern-Simons Lifshitz theory to KdV 
with $m=z, n=N$ can be regarded as a part of the whole map from $hs(N), z$ Chern-Simons Lifshitz theory to KP with $m=z$, with $N$ being the parameter of the map.

In general  it is possible to define powers  of the pseudo differential operator $S$ to for non integer exponents   \cite{FigueroaO'Farrill:1992cv,Khesin:1993ww}.  We conjecture that by choosing  $N$ as a  real number $\lambda$  we will be able to construct a map between Chern-Simons Lifshitz theory with generic 
$hs(\lambda)$ and KP. We leave the explicit construction of this map for future work, but 
observe that there are several arguments that indicate that this correspondence indeed exists. First, finding the maps involves solving  algebraic equations, as in the case of $\lambda=N$, 
but the recursive solution  does   in general not require $N$ to be an integer. Second, the 
$hs(\lambda)$ Chern-Simons for a conformal theory provides a realization of the $W_\infty
$ nonlinear extension of the $W_N$ algebras \cite{Gaberdiel:2012uj,Gaberdiel:2011wb,Campoleoni:2011hg}. While the construction is slightly different many of the 
features of the relation such as the relation of the gauge transformations  which preserve 
the highest weight gauge of $a_c$ to the $W$-algebra transformation,  carry over. When 
$W$ algebras were first investigated in the early '90 a relation of the $W_\infty$ algebra to the 
KP hierarchy was proposed in several papers \cite{FigueroaO'Farrill:1992cv,FigueroaO'Farrill:1992uq,Khesin:1993ww,Khesin:1993ru,Yu:1991bk,Yu:1991ng}

\section{Lifshitz symmetry algebra for generic $hs(\lambda)$ and arbitrary $z$}
\setcounter{equation}{0}
\label{sec7}

In this section we show how Lifshitz symmetry algebra is realized in the generic case, that is, with gauge algebra $hs(\lambda)$ and an arbitrary integer Lifshitz scaling exponent $z$, with special gauge choice. We have the asymptotic Lifshitz connection
\bea
 &&a_x=V^2_1+\sum_{i=2}^\infty \alpha_i V^i_{-i+1} \nonumber\\
 &&a_t=(*a_x)^z+\Delta a_t.
\eea
We choose a slightly different gauge which is called "non highest weight gauge" for $a_t$, that is, the only highest weight term in $a_t$ is $V^{z+1}_z$ (see appendix \ref{appb} for some details on this gauge choice and the nomenclature we are using).

The generic infinitesimal gauge transformation preserving the radial gauge is generated by the gauge parameter $\Lambda(\rho,x,t)=b(\rho)^{-1}\lambda(x,t)b(\rho)$ and the gauge transformation itself is
\be
 \delta a_\mu= [a_\mu,\lambda]+\partial_\mu \lambda
\ee
The three gauge parameters generating time translation, space translation and Lifshitz scaling are
\bea
 \lambda_H&=&-a_t \nonumber\\
 \lambda_P&=&-a_x \nonumber\\
 \lambda_D&=&x a_x+zt a_t-V^2_0,
\eea
We can verify directly by the flatness condition that these gauge parameters generate the desired transformations and preserve lowest weight gauge for $a_x$.
We will use the general formula
\be
 \delta Q(\Lambda)=-{k \over 2 \pi} \int dx {\rm tr}(\Lambda \delta A_x)=-{k \over 2 \pi} \int dx {\rm tr}(\lambda \delta a_x)  \label{BoundaryCharge}
\ee
to obtain the boundary charges.
The Hamiltonian can be expressed as follows:
\bea
 &&\delta Q(\Lambda_H)={k \over 2\pi} \int dx {\rm tr} (a_t \delta a_x)={k \over 2\pi} {\rm tr} (V^{z+1}_z*V^{z+1}_{-z}) \int dx \delta \alpha_{z+1} \nonumber\\
 &&Q(\Lambda_H)={k \over 2\pi} {\rm tr} (V^{z+1}_z*V^{z+1}_{-z}) \int dx \alpha_{z+1} 
\eea
The 
momentum is given by
\bea
 &&\delta Q(\Lambda_P)={k \over 2\pi} \int dx {\rm tr} (a_x \delta a_x)={k \over 2\pi} {\rm tr} (V^2_1*V^2_{-1}) \int dx \delta \alpha_2 \nonumber\\
 &&Q(\Lambda_P)={k \over 2\pi} {\rm tr} (V^2_1*V^2_{-1}) \int dx \alpha_2 
\eea
The Lifshitz scaling charge takes the form
\be
 Q(\Lambda_D)=-{k \over 2\pi} \int dx x\alpha_2 {\rm tr}(V^2_1*V^2_{-1}) + zt\alpha_{z+1} {\rm tr}(V^{z+1}_z*V^{z+1}_{-z})  
\ee
Now let's use the formula \cite{Campoleoni:2011hg}
\be
 \{Q(\Lambda),Q(\Gamma)\}=\delta_\Lambda Q(\Gamma)=-\delta_\Gamma Q(\Lambda) \label{Bracket}
\ee
to verify the Lifshitz algebra.
\bea
 \{Q(\Lambda_H),Q(\Lambda_P)\}
 &=&-\delta_{\Lambda_P} Q(\Lambda_H)={k \over 2\pi} {\rm tr} (V^{z+1}_z*V^{z+1}_{-z}) \int dx \alpha_{z+1}^{'}=0  
\eea
with the by-product that $\dot{\alpha_2}$ must be a total derivative.
\be
 \{Q(\Lambda_D),Q(\Lambda_H)\}=\delta_{\Lambda_D} Q(\Lambda_H)={k \over 2\pi} {\rm tr} (V^{z+1}_z*V^{z+1}_{-z}) \int dx \delta_{\Lambda_D} \alpha_{z+1} 
\ee
using
\bea
 &&\delta_{\Lambda_D} a_x=\partial_x \lambda_D + [a_x,\lambda_D]=a_x+[V^2_0,a_x]+x a_x^{'}+zt \dot{a_x} \nonumber\\
 &&\delta_{\Lambda_D} \alpha_{z+1} = (z+1) \alpha_{z+1} + x \alpha_{z+1}^{'} + zt \dot{\alpha_{z+1}} 
\eea
we get
\be
 \{Q(\Lambda_D),Q(\Lambda_H)\}=z Q(\Lambda_H).
\ee
At last
\be
 \{Q(\Lambda_D),Q(\Lambda_P)\}=\delta_{\Lambda_D} Q(\Lambda_P)={k \over 2\pi} {\rm tr} (V^2_1*V^2_{-1}) \int dx \delta_{\Lambda_D} \alpha_2
\ee
using
\be
 \delta_{\Lambda_D} \alpha_2 = 2 \alpha_2 + x \alpha_2^{'} + zt \dot{\alpha_2}
\ee
we get
\be
 \{Q(\Lambda_D),Q(\Lambda_P)\}=Q(\Lambda_P).
\ee

\subsection{The $sl(3,\reals)$ and $z=2$ case}
In this section with $sl(3,\reals)$, $z=2$, we will obtain boundary charges corresponding to time translation, space translation and Lifshitz scaling and that verify they satisfy Lifshitz algebra. Furthermore, we identify the components of the stress energy tensor and show it's consistent with conservation laws and Lifshitz scaling symmetry.

For $sl(3,\reals)$ The generic gauge parameter is
\be
\lambda=\sum_{i=-1}^1 \epsilon_i V^2_i + \sum_{j=-2}^2 \chi_j V^3_j.
\ee
By requiring $a_x$ to be form-invariant, that is, it's still in the lowest weight gauge, we find only the coefficient of highest weight terms $\epsilon_1, \chi_2$ are free and all the other variables in the gauge parameter are expressed in terms of them.
We can assign specific values to $\epsilon_1, \chi_2$ to get gauge parameters generating time translation, space translation and Lifshitz scaling
\bea
&&\epsilon_1=0, \chi_2=-1, \lambda=\lambda_H \nonumber\\
&&\epsilon_1=-1, \chi_2=0, \lambda=\lambda_P \nonumber\\
&&\epsilon_1=x, \chi_2=2t, \lambda=\lambda_D
\eea
Using (\ref{BoundaryCharge}) we get the symmetry charges
\bea
Q(\Lambda_H)&=& {2k \over \pi} \int dx \alpha_3 \nonumber\\
Q(\Lambda_P)&=& -{2k \over \pi} \int dx \alpha_2 \nonumber\\
Q(\Lambda_D)&=& {2k \over \pi} \int dx ( x \alpha_2 - 2t \alpha_3)
\eea
Using (\ref{Bracket}) we can verify the Lifshitz algebra
\bea
  \{Q(\Lambda_H),Q(\Lambda_P)\}&=&0 \nonumber\\
  \{Q(\Lambda_D),Q(\Lambda_H)\}&=&2Q(\Lambda_H) \nonumber\\
  \{Q(\Lambda_D),Q(\Lambda_P)\}&=&Q(\Lambda_P).
\eea
Identify the density of $Q(\Lambda_H)$ as the energy density, the density of $Q(\Lambda_P)$ as the momentum density
\bea
 {\cal E}&=& {2k \over \pi} \alpha_3 \nonumber\\
 {\cal P}_x&=&-{2k \over \pi} \alpha_2.
\eea
Use the Lifshitz symmetry condition ${2 \cal E}+\Pi^x_x=0$ to get $\Pi^x_x=-{4k \over \pi} \alpha_3$, we can verify that the conservation of momentum
\be
 \partial_t {\cal P}_x + \partial_x \Pi^x_x = 0
\ee
is guaranteed by the equations of motion.
 Plugging  the expression for $\cal E$ into the equation of conservation of energy $\partial_t {\cal P} + \partial_x {\cal E}^x=0$ one obtains  the expression for energy flow 
 \be
 {\cal E}^x=-{2k \over \pi}(\frac{2}{3} \alpha_2^2 + \frac{1}{6} \alpha_2^{''})
 \ee

\subsection{The $sl(4,\reals)$ and $z=3$ case}

Analogues to what we did for $sl(3,\reals)$, $z=2$, we will obtain boundary charges, verify the symmetry algebra and study the stress energy tensor. Here we work with KdV gauge.
In $sl(4,\reals)$, the generic gauge parameter for infinitesimal gauge transformation is
\be
\lambda=\sum_{i=-1}^1 \epsilon_i V^2_i + \sum_{j=-2}^2 \chi_j V^3_j + \sum_{k=-3}^3 \mu_k V^4_k.
\ee
By requiring $a_x$ to be form-invariant, we find only the highest weight terms $\epsilon_1, \chi_2, \mu_3$ are free. Again by appropriately choosing values for these three variables, we get the desired gauge parameters
\bea
&&\epsilon_1={7 \over 10}\alpha_2, \chi_2=0, \mu_3=-1,\lambda=\lambda_H \nonumber\\
&&\epsilon_1=-1, \chi_2=0, \mu_3=0, \lambda=\lambda_P \nonumber\\
&&\epsilon_1=x-{21 \over 10}\alpha_2 t, \chi_2=0, \mu_3=3t, \lambda=\lambda_D
\eea
For these three gauge parameters, we use (\ref{BoundaryCharge}) to calculate the boundary charges
\bea
Q(\Lambda_H)&=& {k \over {2\pi}} \int dx (-36\alpha_4+ \dfrac{7}{2} \alpha_2^2) \nonumber\\
Q(\Lambda_P)&=& {k \over {2\pi}} \int dx (-10 \alpha_2) \nonumber\\
Q(\Lambda_D)&=& {k \over {2\pi}} \int dx (10 x \alpha_2 + 108t \alpha_4 + 21t \alpha_2^2)
\eea
Again use (\ref{Bracket}) we can verify the Lifshitz symmetry algebra.
The density of $Q(\Lambda_H)$ is identified with  the energy density up to a total derivative and the density of $Q(\Lambda_P)$ is identified with the momentum density
\bea
 {\cal E}&=& {k \over {2\pi}}(-36\alpha_4+ \dfrac{7}{2} \alpha_2^2 + \frac{7}{6} \alpha_2^{''}) \nonumber\\
 {\cal P}_x&=& -{k \over {2\pi}} 10 \alpha_2
\eea
Using  the Lifshitz symmetry condition ${3 \cal E}+\Pi^x_x=0$ to get $\Pi^x_x=-3 \cal E$, we can show that
\be
\partial_t {\cal P}_x+\partial_x \Pi^x_x=0.
\ee
The relation of the KdV conserved charges $q^{(i)}$    to the Chern-Simons variables is given by the following expressions 
\bea
  &&q^{(1)}=\int dx \frac{1}{4} u_2=\int dx \frac{5}{2} \alpha_2 \nonumber\\
  &&q^{(2)}=\int dx \frac{1}{4} u_3=\int dx\, 6 \alpha_3 \nonumber\\
  &&q^{(3)}=\int dx (\frac{3}{4}u_4+ \frac{1}{16} u_2^{''} - \frac{3}{32} u_2^2 - \frac{3}{8} u_3^{'})=\int dx (27\,\alpha_4-\frac{21}{8} \alpha_2^2) \nonumber
\eea
Hence, the KdV charges  $q^{(1)}$ and $q^{(3)}$ are proportional to  $Q(\Lambda_P)$  and $Q(\Lambda_H)$ respectively.

\section{Discussion}
\label{sec8}

In the present paper we have explored the relation of asymptotic Lifshitz spacetimes in  higher spin 
gravity theories to  integrable systems in the KdV hierarchy. We were able to make this relation 
explicit in some specific examples. The evidence of the validity of the general conjecture is the  
match of the number of degrees of freedom, the agreement of the scaling symmetry of the fields on 
both sides and the presence of a residual gauge symmetry which can be used to construct the 
exact map of the equations. It's an interesting open question to find a general proof for our conjecture and we hope to come back to this question  in future work.

The Chern-Simons formulation of higher spin gravity and $W$-algebras  are strongly related. $W$-algebras are  nonlinear extensions the 2-dimensional conformal algebra with higher spin fields. For 
example the standard Drinfeld-Sokolov reduction relates the symmetries of the asymptotically 
conformal $sl(N,\reals)$  connections to the symmetries and Ward identities of the $W_N$ 
algebra.  In the past there have been several generalizations of the  result relating the Virasoro 
algebra to the KdV equation, relating the elements of the KdV hierarchy to $W$ algebras in the 
context of conformal field theories and their deformations \cite{Zamolodchikov:1987jf,Mathieu:1991et,Kupershmidt:1989bf}.

In the present paper we constructed Chern-Simons connections which do not have a conformal 
scaling symmetry but instead a time/space anisotropic Lifshitz scaling symmetry. It is an 
interesting question how our  results are related to above mentioned work on the relation of $W$-algebras and KdV. 
One important difference in the construction of the connections  is that we do not use   light-cone 
coordinates $x^\pm$ (or complex coordinates $z,\bar z$ after Wick rotation). Light cone coordinates naturally lead  to 
split into left and right movers (or holomorphic and anti-holomorphic) sectors characteristic of a CFT. 
On the CFT side the translations in both space and time are generated by modes of the stress tensor.

 In the construction of the Lifshitz theories we use time $t$ and a spatial coordinate $x$ instead. The analysis of section \ref{sec7} shows   that spatial translations are generated by 
 a charge associated with the stress tensor, whereas time translations are generated by a charge associated with the higher spin current in the $W_N$-algebra. This suggest an intriguing 
 possibility to construct a Lifshitz theory with scaling exponent $z=N$ from a $W_N$ CFT: define the theory on a spatial slice and replace the Hamiltonian which generates time evolution by the integrated spin $N$ current.  We leave the exploration of this suggestion for future work.

\section*{Acknowledgements}

It is a pleasure to thank  Eliot Hijano and Josh Samani for useful discussion.  This work was supported in part by National Science Foundation grant PHY-13-13986

\newpage

\appendix
\section{$sl(3,\reals)$, $sl(4,\reals)$ and $hs(\lambda)$ conventions}
\setcounter{equation}{0}
\label{appa}

In this appendix we present a realization of the $sl(N,\reals)$ algebra which are used for calculations in the main body of the text.
\subsection{$sl(3,\reals)$ }
The $sl(2,\reals)$  generators of the principal embedding  are given by the
following matrices
\begin{align}
  L_{-1} = \begin{pmatrix}
          0 & \sqrt 2 & 0 \\
          0 & 0 & \sqrt 2 \\
          0 & 0 & 0 \\
        \end{pmatrix}, \qquad
  L_1 = \begin{pmatrix}
          0 & 0 & 0 \\
          -\sqrt 2 & 0 & 0 \\
          0 & -\sqrt 2 & 0 \\
        \end{pmatrix}, \qquad
  L_0 = \begin{pmatrix}
          1 & 0 & 0 \\
          0  & 0 & 0 \\
          0 & 0  & -1 \\
        \end{pmatrix}
\end{align}
and the spin 3 generators, on which we omit the superscript $^{(3)}$ for
notational simplicity, are
as
follows:
\begin{align}
  W_{-2} &= \begin{pmatrix}
          0 & 0 & 2 \\
          0  & 0 & 0 \\
          0 & 0  & 0 \\
        \end{pmatrix}, \qquad
  W_{-1} = \begin{pmatrix}
          0 & \frac{1}{\sqrt 2} & 0 \\
          0 & 0 & -\frac{1}{\sqrt 2}\\
          0 & 0 & 0 \\
        \end{pmatrix}, \qquad
  W_0 = \begin{pmatrix}
          \frac{1}{3} & 0 & 0 \\
          0  & -\frac{2}{3} & 0 \\
          0 & 0  & \frac{1}{3} \\
        \end{pmatrix} \\
  W_1 &= \begin{pmatrix}
          0 & 0 & 0 \\
          -\frac{1}{\sqrt 2} & 0 & 0 \\
          0 & \frac{1}{\sqrt 2} & 0 \\
        \end{pmatrix}, \qquad
  W_2 = \begin{pmatrix}
          0 & 0 & 0 \\
          0  & 0 & 0 \\
          2 & 0  & 0 \\
        \end{pmatrix}
\end{align}
If we define $(T_1,T_2, \dots, T_8) = (L_1, L_0, L_{-1}, W_2, \dots W_{-2})$, then
traces of all
pairs
of generators are given by
\begin{align}\label{trsl3}
  \mathrm{tr}(T_iT_j)
  &=
  \left(\begin{array}{ccc|ccccc}
        &   & -4  &   0  &    &        \cdots       &       &0\\
            & 2   &      &  \vdots  &    &    \ddots           &       &\vdots\\
    -4  &   &     &    0  &    &        \cdots       &       &0\\
       \hline
         0   & \cdots  &  0   &     &   &               &       & 4\\
            &   &     &     &   &               & -1    &\\
         \vdots   & \ddots  &  \vdots   &     &   & \frac{2}{3}   &       &\\
            &   &     &     & -1&               &       &\\
          0  & \cdots  &   0  & 4   &   &               &       &
     \end{array}\right)
\end{align}

\subsection{$sl(4,\reals)$ }
The $sl(4,\reals)$ matrix representation we use is the following. The   $sl(2,\reals)$ sub algebra given by
\bea
&&l_0=\left(
\begin{array}{cccc}
 -\frac{3}{2} & 0 & 0 & 0 \\
 0 & -\frac{1}{2} & 0 & 0 \\
 0 & 0 & \frac{1}{2} & 0 \\
 0 & 0 & 0 & \frac{3}{2} \\
\end{array}
\right)\quad  l_1=\left(
\begin{array}{cccc}
 0 & 1 & 0 & 0 \\
 0 & 0 & 1 & 0 \\
 0 & 0 & 0 & 1 \\
 0 & 0 & 0 & 0 \\
\end{array}
\right)\quad l_{-1}=\left(
\begin{array}{cccc}
 0 & 0 & 0 & 0 \\
 -3 & 0 & 0 & 0 \\
 0 & -4 & 0 & 0 \\
 0 & 0 & -3 & 0 \\
\end{array}
\right)\nonumber\\
\eea
 $w_i, i=+2,+1,\cdots,-2$ form a spin 2 representation, whereas the $u_i, i=+3,+3,\cdots,-3$ form  s spin 3 representation of the $sl(2,\reals)$ sub algebra.
\bea
&& w_2=\left(
\begin{array}{cccc}
 0 & 0 & 1 & 0 \\
 0 & 0 & 0 & 1 \\
 0 & 0 & 0 & 0 \\
 0 & 0 & 0 & 0 \\
\end{array}
\right)\quad w_1=\left(
\begin{array}{cccc}
 0 & -1 & 0 & 0 \\
 0 & 0 & 0 & 0 \\
 0 & 0 & 0 & 1 \\
 0 & 0 & 0 & 0 \\
\end{array}
\right)\quad w_0=\left(
\begin{array}{cccc}
 1 & 0 & 0 & 0 \\
 0 & -1 & 0 & 0 \\
 0 & 0 & -1 & 0 \\
 0 & 0 & 0 & 1 \\
\end{array}
\right)\nonumber\\
&&w_{-1}= \left(
\begin{array}{cccc}
 0 & 0 & 0 & 0 \\
 3 & 0 & 0 & 0 \\
 0 & 0 & 0 & 0 \\
 0 & 0 & -3 & 0 \\
\end{array}
\right) \quad  w_{-2}=\left(
\begin{array}{cccc}
 0 & 0 & 0 & 0 \\
 0 & 0 & 0 & 0 \\
 12 & 0 & 0 & 0 \\
 0 & 12 & 0 & 0 \\
\end{array}
\right)\quad u_3=\left(
\begin{array}{cccc}
 0 & 0 & 0 & 1 \\
 0 & 0 & 0 & 0 \\
 0 & 0 & 0 & 0 \\
 0 & 0 & 0 & 0 \\
\end{array}
\right)\nonumber\\
&&u_2=\left(
\begin{array}{cccc}
 0 & 0 & -\frac{1}{2} & 0 \\
 0 & 0 & 0 & \frac{1}{2} \\
 0 & 0 & 0 & 0 \\
 0 & 0 & 0 & 0 \\
\end{array}
\right)\quad u_1=\left(
\begin{array}{cccc}
 0 & \frac{2}{5} & 0 & 0 \\
 0 & 0 & -\frac{3}{5} & 0 \\
 0 & 0 & 0 & \frac{2}{5} \\
 0 & 0 & 0 & 0 \\
\end{array}
\right)u_0=\left(
\begin{array}{cccc}
 -\frac{3}{10} & 0 & 0 & 0 \\
 0 & \frac{9}{10} & 0 & 0 \\
 0 & 0 & -\frac{9}{10} & 0 \\
 0 & 0 & 0 & \frac{3}{10} \\
\end{array}
\right)\nonumber\\
&&u_{-1}=\left(
\begin{array}{cccc}
 0 & 0 & 0 & 0 \\
 -\frac{6}{5} & 0 & 0 & 0 \\
 0 & \frac{12}{5} & 0 & 0 \\
 0 & 0 & -\frac{6}{5} & 0 \\
\end{array}
\right)\quad u_{-2}=\left(
\begin{array}{cccc}
 0 & 0 & 0 & 0 \\
 0 & 0 & 0 & 0 \\
 -6 & 0 & 0 & 0 \\
 0 & 6 & 0 & 0 \\
\end{array}
\right)\quad u_{-3}=\left(
\begin{array}{cccc}
 0 & 0 & 0 & 0 \\
 0 & 0 & 0 & 0 \\
 0 & 0 & 0 & 0 \\
 -36 & 0 & 0 & 0 \\
\end{array}
\right)\nonumber\\
\eea
The $w_i, i=+2,+1,\cdots,-2$ form a spin 2 representation, whereas the $u_i, i=+3,+3,\cdots,-3$ form  s spin 3 representation of the $sl(2,\reals)$ sub algebra.

\subsection{$hs(\lambda)$ conventions}
\label{hslamconv}
Higher spin algebra elements $V^s_m$, $s=1,2,3,\ldots$ and $m=-s+1,-s+2,\ldots,s-1$. We call $s$ the spin and $m$ the weight.

The lone star product is defined as
\be
 V^s_m * V^t_n = \frac{1}{2} \sum_{u=1}^{s+t-|s-t|-1} g^{st}_u(m,n,\lambda) V^{s+t-u}_{m+n}
\ee
The structure constants of the $hs(\lambda)$ algebra were defined in
\cite{Pope:1989sr} and can be represented as follows
\be\label{gdefa}
g_u^{st}(m,n;\lambda) = {q^{u-2}\over2(u-1)!}\phi_u^{st}(\lambda)N_u^{st}(m,n)
\ee
$q$ is a normalization constant which can be eliminated by a rescaling on the generators, we choose $q=1/4$ to agree with the literature. The other terms in (\ref{gdefa}) are given by 
\bea
 N_u^{st}(m,n) &=& \sum_{k=0}^{u-1}(-1)^k
\left( 
\begin{array}{c}
     u-1  \\
k
\end{array}
\right)
[s-1+m]_{u-1-k}[s-1-m]_k[t-1+n]_k[t-1-n]_{u-1-k}\nonumber\\
\phi_u^{st}(\lambda) &= &\ _4F_3\left[
\begin{array}{cccc}
 \half + \lambda  &   \half - \lambda &{2-u\over 2}   &{1-u\over 2}\\
 {3\over 2}-s  &  {3\over 2} -t &  \half + s+t-u &\\  
\end{array} \Bigg| 1
\right]
\eea
The descending Pochhammer symbol $[a]_n$ is defined as,
\be
[a]_n  = a(a-1)...(a-n+1)~
\ee
The commutator is defined as
\be
 [V^s_m, V^t_n]=V^s_m * V^t_n - V^t_n * V^s_m
\ee
$V^1_0$ is the unit element. The trace of a $hs(\lambda)$ element is defined as the coefficient of $V^1_0$ up to a multiplicative constant ${\rm tr}(V^1_0)$. When $\lambda=N$ where $N$ is a positive integer, $hs(\lambda)$ is truncated to $sl(N,\reals)$. That means, we can consistently set $V^s_m$ to be zero if $s>N$, and the remaining elements form $sl(N,\reals)$ with star product identified as matrix multiplication and trace identified as matrix trace.

\section{Gauge choices}
\setcounter{equation}{0}
\label{appb}

\subsection{Lowest weight gauge}
The lowest weight gauge is that $a_x$ only contains lowest weight terms except for $V^2_1$. Here we show how we can transform away all non lowest weight terms in $a_x$.
Under an infinitesimal gauge transformation
\be
 \delta a_x=[a_x,\lambda]+\partial_x \lambda
\ee
We have $V^2_1$ in $a_x$, so we can put $V^s_{s-2}$ in the gauge parameter $\lambda$ to gain a highest weight term $V^s_{s-1}$ in $\delta a_x$ from the commutator. We can exponentiate this infinitesimal transformation to cancel the highest weight term in the original $a_x$. After eliminating all highest weight terms, we use $V^s_{s-3}$ in $\lambda$ to cancel $V^s_{s-2}$ terms. Do this recursively we get to the lowest weight gauge.

\subsection{Gauge freedom of $a_t$ and non highest weight gauge}
The construction of the asymptotic Lifshitz connection used  the fact that $-\partial_x a_t+[a_t,a_x]$ should only contain lowest weight terms to determine coefficients of $a_t$ in terms of coefficients of $a_x$ up to some indeterminacy. The computation is the same as to find an infinitesimal gauge transformation preserving lowest weight gauge, with $-a_t$ playing the role of the gauge parameter. Therefore the indeterminacy in $a_t$ is a  gauge freedom. The indeterminacy encompasses actually all the coefficients of the highest weight terms in $\Delta a_t$. We can choose them to make highest weight terms in $a_t$ to vanish except for the leading term. We call it non highest weight gauge for $a_t$.
 
\newpage

\end{document}